\documentclass[12pt]{article} 
\usepackage{amssymb}
\usepackage{graphicx}
\begin{document} 
\begin{center}
{\large \bf Proton structure and elastic scattering amplitudes}

\vspace{0.5cm}                   

{\bf I.M. Dremin}

\vspace{0.5cm}                       

         Lebedev Physical Institute, Moscow 119991, Russia\\

\end{center}

\begin{abstract}
Three main statements are advocated in this talk:

1. Protons become more active at the periphery with increase of
their collision energy as follows from comparison of ISR and LHC data.

2. The geometric scaling is violated even in the diffraction region
as follows from comparison of lower energy and LHC data.

3. The problem of the ratio of real to imaginary parts of the elastic
scattering amplitude at non-zero transferred momenta is very crucial.

The talk is based on arXiv:1206.5474 (review), 1202.2016, 1204.1914,
1204.4866, 1208.3073, 1209.1935, 1212.3313, 1304.5345, 1306.5384.
All papers have been published already.
\end{abstract}

1. More details are in hep-ph:1306.5384.

Protons become larger and more black with energy increase from ISR to LHC
energies. This follows from the impact parameter analysis of the unitarity
condition for the elastic scattering amplitude $f(s,t)$ which is
\begin{equation}
2{\rm Re}\Gamma (s,b)=\vert \Gamma (s,b)\vert ^2+G(s,b),
\label{unit}
\end{equation}
where $i\Gamma (s,b)$ is the Fourier transform of $f(s,t)$,
$G$ is the overlap function. The unitarity condition is the 
rigorous relation of the Nature which states that the total probability
of outcomes of any particle collision sums to 1.

The smallness of the real part of $f(s,t)$ corresponding to small
${\rm Im}\Gamma (s,b)$ implies that one can compute $G$ approximately as
\begin{equation}
G(s,b)\approx 2{\rm Re}\Gamma (s,b)-({\rm Re}\Gamma (s,b))^2.
\label{over}
\end{equation}
The overlap function $G$ describes the particle 
distribution $d\sigma /db$ in the impact parameter space. One may treat it
as a parton distribution if one-to-one correspondence of particles and partons 
is assumed. It was computed from ISR and LHC data and the difference between
the two shapes is shown in Fig. 1. 
\begin{figure}
\includegraphics[width=\textwidth, height=6.2cm]{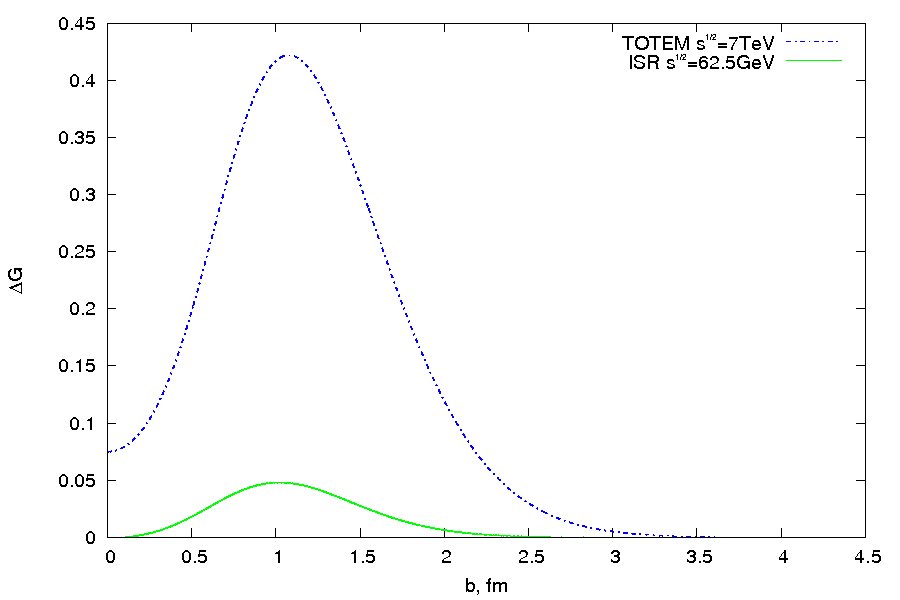}

Fig. 1. The difference between the overlap functions. Dash-dotted curve is for 
7 TeV and 23.5 GeV energies, solid curve is for 62.5 GeV and 23.5 GeV energies.
Conclusion: The parton density at the periphery increases strongly!
\end{figure}
It is concentrated at the periphery of the proton at the distance about 1 fm. 
The increase amounts to about 40$\%$!
Thus, at higher energies, the periphery becomes darker and more populated by 
partons which play more active role in particle production. I ascribe the 
peripheral nature of this effect to strong increase of the cross section of 
the inelastic diffraction with large masses and high multiplicities, which can 
hardly be separated by the gap criteria from the minimum bias events. Protons 
become more black also in the central region at $b$=0 at the level of 8$\%$ 
which does not violate the unitarity bound $G(s,b)\leq 1$. No increase is seen
in Fig. 1 at ISR energies.

2. More details are in hep-ph:1212.3313, 1209.1935.

The geometric scaling of elastic scattering amplitudes was initially proposed
by an analogy to the KNO-scaling of multiplicity distributions in inelastic 
processes. Here, it was supposed that there exists an universal dependence of 
distributions of $t^2d\sigma /dt$ (or equivalently of $\sigma _t^{-2}d\sigma /dt$)
at different energies on the product of the transferred momentum 
and the total cross section $t\sigma _t$. This assumption was approximately 
supported by experimental data up to ISR energies within the diffraction cone.
Moreover it was shown in the above papers that such behavior follows if one
equates the local dispersion expression for the ratio $\rho (s,t)$ which 
contains the $s$-derivative and Martin formula which describes its 
$t$-evolution. However the comparison with LHC data demonstrates the violation
of the geometric scaling at these energies as seen in Fig. 2. The approximate
scaling can be restored if the dependence of $t^{2a}d\sigma /dt$ on 
$t^a\sigma _t$ with $a\approx 1.2$ is plotted. The parameter $a$ is directly
connected with different energy behavior of the total cross section and the
slope of the diffraction cone. Martin formula must be modified.
\begin{figure}
 \includegraphics[width=\textwidth, height=6.2cm]
{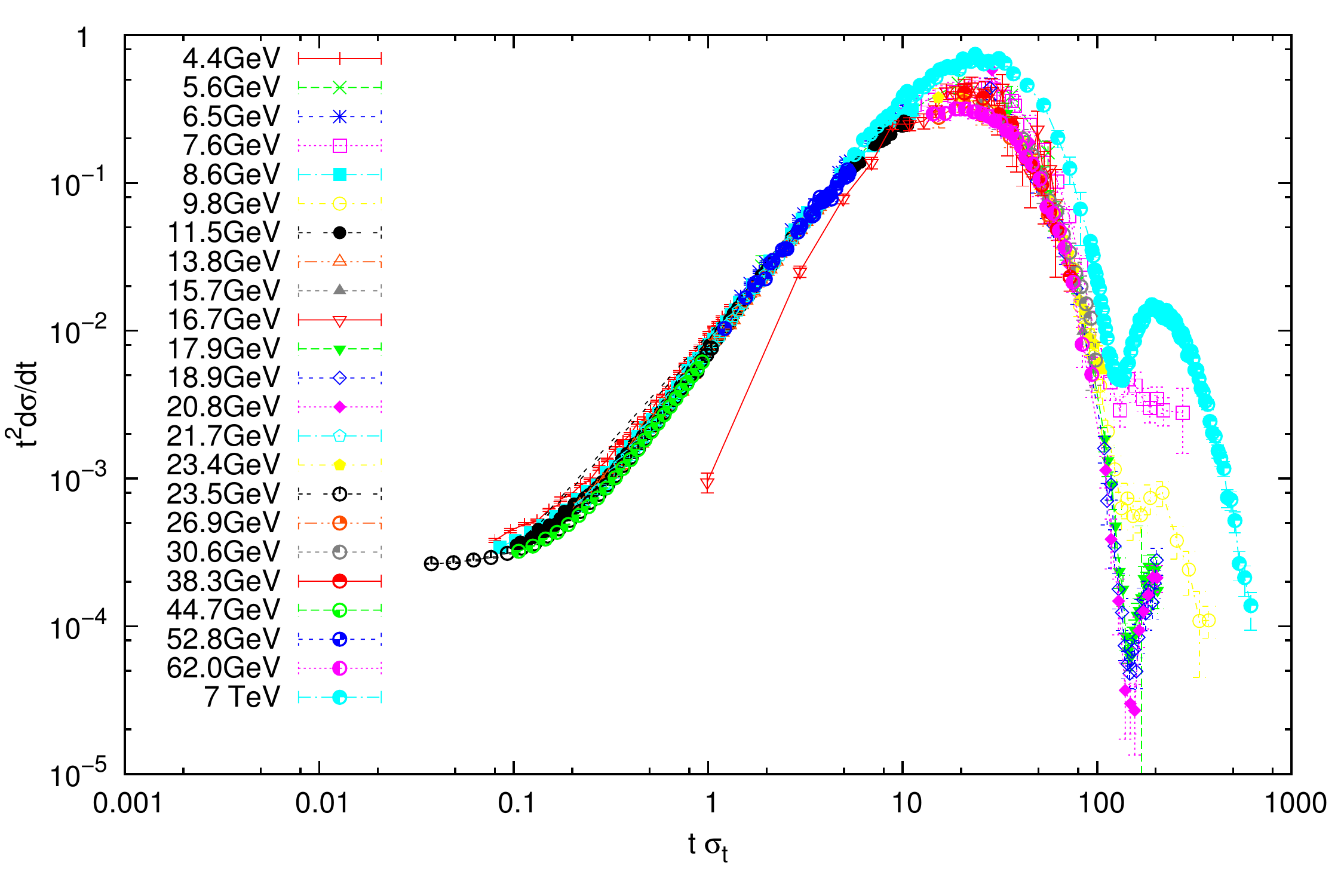}
Fig. 2. The values of $t^2d\sigma /dt$ for pp-scattering at energies $\sqrt s$
from 4.4 GeV to 7 TeV as functions of $t\sigma _t$ with $\sigma _t$ 
of the corresponding experiment.
\end{figure}

3. More details are in hep-ph:1202.2016, 1204.1914, 1204.4866, 1208.3073, 
1304.5345.

At t = 0, the ratio $\rho (s,t)$ is known from Coulomb-nuclear interference
experimentally and from dispersion relations theoretically. The only approach
to non-zero transferred momenta proposed up to now is based on the predictions
for the behavior of $d\sigma /dt$ in the Orear region obtained from the 
unitarity condition. This prediction prescribes $\exp (-r\sqrt {\vert t\vert })$ 
decrease in there with the well defined exponential $r$ depending on 
$\rho (s,t)$ and very sensitive to it. The comparison with LHC data has
shown that this ratio must be negative and quite large (about -2) in this
region. Most of the widely used models do not predict such values. Moreover
many of them get it positive. This follows from the equal numbers of zeros of 
real and imaginary parts. Only those models with uneven sum of this number
can succeed in getting negative $\rho (s,t)$. 
The unitarity condition does not ask for a zero of the imaginary part to fit
the dip as the models do but ascribes it to the damped oscillations contained
in the solution of the equation. The difference of the 
models and the unitarity condition is not resolved yet. 
The general principle "Each talk must contain one and only one statement" can be 
restored if one accepts the above statements as three separate talks.
This work is partially supported by RFBR and by WP8 of the hadron physics 
program of the 8th EU program period.
\end{document}